\documentclass[10pt,letterpaper,twocolumn]{article}
\usepackage[top=0.85in,left=1.0in,right=1.0in,footskip=0.75in,marginparwidth=2in]{geometry}

\usepackage{mathtools}  
\usepackage{amssymb}
\usepackage[T1]{fontenc}
\usepackage{ae,aecompl}
\usepackage{adjustbox}

\usepackage{graphicx}	
\usepackage{amsmath}	
\usepackage{amssymb}	
\usepackage[numbers]{natbib} 
\usepackage{hyperref}
\usepackage{fancyhdr}

\begin{document}

\title{\textbf{Stationary waves and slowly moving features in the night upper clouds of Venus}}


\author{J. Peralta$^1$\thanks{E-mail: javier.peralta@ac.jaxa.jp}\footnote{$^1$Institute of Space and Astronautical Science (ISAS), Japan Aerospace Exploration Agency (JAXA), Japan $^2$Grupo de Ciencias Planetarias, Departamento de F{\'i}sica Aplicada I, E.T.S. Ingenier{\'i}a, Universidad del Pa{\'i}s Vasco (UPV/EHU), Spain $^3$Zentrum f{\"u}r Astronomie und Astrophysik, Technische Universit{\"a}t Berlin, Germany $^4$Artificial Intelligence Research Center, National Institute of Advanced Industrial Science and Technology, Japan $^5$Faculty of Science, Kyoto Sangyo University, Japan $^6$Istituto di Astrofisica e Planetologia Spaziali (INAF-IAPS), Italy $^7$Abteilung Planetenforschung, Rheinisches Institut f{\"u}r Umweltforschung, Universit{\"a}t zu K{\"o}ln, Cologne, Germany $^8$Graduate School of Frontier Sciences, University of Tokyo, Japan.}, R. Hueso$^2$, A. S{\'a}nchez-Lavega$^2$, Y.~J. Lee$^1$, A. Garc{\'i}a-Mu{\~n}oz$^3$, T. Kouyama$^4$,\\
H. Sagawa$^5$, T.~M. Sato$^1$, G. Piccioni$^6$, S. Tellmann$^7$, T. Imamura$^8$ \& T. Satoh$^1$}

\date{Published in \textit{Nature Astronomy} the 24 July 2017; original manuscript \href{https://www.nature.com/articles/s41550-017-0187}{here}.}


\maketitle
\thispagestyle{fancy}


\textbf{At the cloud top level of Venus (65-70 km altitude) the atmosphere rotates 60 times faster than the underlying surface, a phenomenon known as superrotation \citep{Gierasch1997,Lebonnois2013}.  Whereas on Venus's dayside the cloud top motions are well determined \cite{Peralta2007b,Kouyama2013,Khatuntsev2013,Hueso2015} and Venus general circulation models predict a mean zonal flow at the upper clouds similar on both day and nightside \citep{Lebonnois2013}, the nightside circulation remains poorly studied except for the polar region \citep{Luz2011,Garate-Lopez2013}. Here we report global measurements of the nightside circulation at the upper cloud level. We tracked individual features in thermal emission images at 3.8 and 5.0 $\mathrm{\mu m}$ obtained between 2006 and 2008 by the Visible and Infrared Thermal Imaging Spectrometer (VIRTIS-M) onboard Venus Express and in 2015 by ground-based measurements with the Medium-Resolution 0.8-5.5 Micron Spectrograph and Imager (SpeX) at the National Aeronautics and Space Administration Infrared Telescope Facility (NASA/IRTF). The zonal motions range from -110 to -60 m s$^{-1}$, consistent with those found for the dayside but with larger dispersion \citep{Hueso2015}. Slow motions (-50 to -20 m s$^{-1}$) were also found and remain unexplained. In addition, abundant stationary wave patterns with zonal speeds from -10 to +10 m s$^{-1}$ dominate the night upper clouds and concentrate over the regions of higher surface elevation.}\\

\null
Nocturnal images taken at wavelengths of about 3.8 and 5.0 $\mathrm{\mu m}$ sense the thermal contrasts (2--30\% in terms of radiance) of the upper clouds at around 65 km above the surface \citep{GarciaMunoz2013,Carlson1991,Drossart2007}. To study the motions of the night clouds of Venus, we used images at these two wavelengths obtained by the Visible and Infrared Thermal Imaging Spectrometer-Mapper (VIRTIS-M) onboard Venus Express (VEx) \citep{Drossart2007}, a mapping spectrometer with moderate spectral resolution (R$\sim$200) which acquired infrared images from June 2006 to August 2008, and ground-based observations at 4.7 $\mathrm{\mu m}$ taken with the Medium-Resolution 0.8-5.5 Micron Spectrograph and Imager (SpeX) at the 3-m NASA's Infrared Telescope Facility (IRTF) in July 2015. The spatial resolution of the images ranged from about 10 km pixel$^{-1}$ for VIRTIS-M to 400 km pixel$^{-1}$ in the case of SpeX (see Methods). Observational constraints due to the polar orbit of VEx limited the number of available images showing mid-to-low latitudes. Additional constraints due to the combination of exposure time and detector temperature reduced our final dataset to 55 VEx orbits ($\sim$55 Earth days), which was useful for nightside feature tracking. SpeX observations extended coverage to the northern hemisphere. Similar thermal images of Venus's south polar region have been used to study the dynamics of the south polar vortex \citep{Luz2011,Garate-Lopez2013}. We improved the contrast of individual images by selecting and averaging the best-quality images (with fewer acquisition defects) within the spectral ranges 3.68--3.94 and 5.00--5.12  $\mathrm{\mu m}$. The SpeX images were obtained with an M'-band filter (4.57--4.79 $\mathrm{\mu m}$). In both cases, we used high-pass filtering to improve the visual contrast of faint features in the images and map projected the images for accurate feature tracking (see Methods).\\

\begin{figure}[h]
\centerline{\includegraphics[width=20pc]{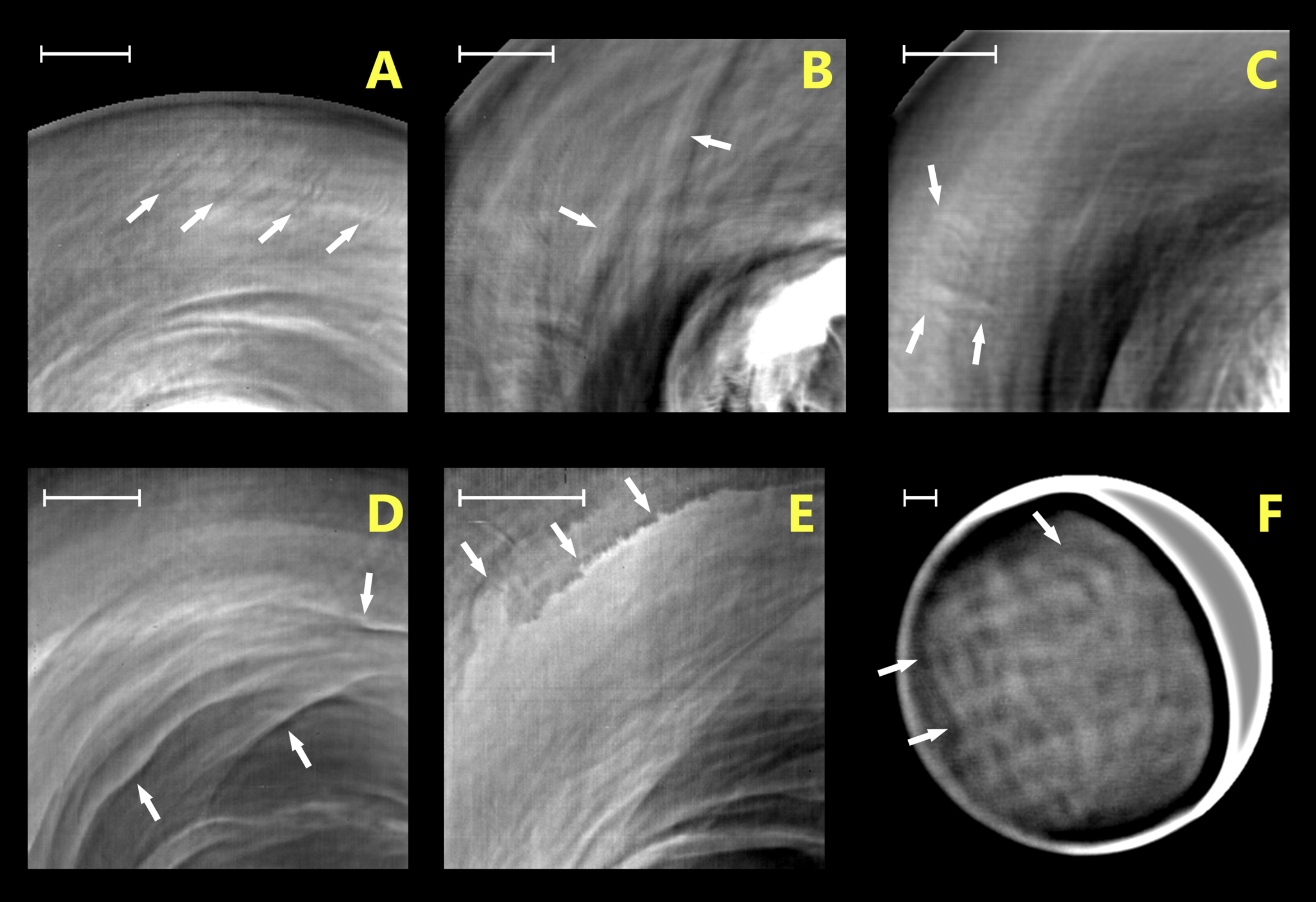}}
\caption{\textbf{Nightside thermal features on the upper clouds of Venus. a}, Wavy patterns (arrows) in a 3.8 $\mathrm{\mu m}$ image at latitudes and longitudes of 30--50$^{\circ}$S and 330--20$^{\circ}$, respectively (VEx orbit 259). \textbf{b}, Long and broad stripes (probably waves; arrows) observed in a 3.8 $\mathrm{\mu m}$ image extending from latitudes of 65$^{\circ}$S to (at least) 40$^{\circ}$S and longitudes of 65--115$^{\circ}$ (VEx orbit 740). \textbf{c}, Further wavy patterns (arrows) in a 5.0 $\mathrm{\mu m}$ image at coordinates 40--60$^{\circ}$S and 10--30$^{\circ}$ (VEx orbit 726). \textbf{d}, Bright filaments (arrows) in a 3.8 $\mathrm{\mu m}$ image at 40--65$^{\circ}$S and 270--330$^{\circ}$ (VEx orbit 594). \textbf{e}, Shear-like structures (arrows) in a 3.8 $\mathrm{\mu m}$ image at 35--50$^{\circ}$S and 290--340$^{\circ}$ with smaller stationary waves at about 300--303$^{\circ}$ and 35--45$^{\circ}$S (VEx orbit 598). \textbf{f}, Big-scale features (arrows) observed in a 5 $\mathrm{\mu m}$ image. Images \textbf{a-e} were obtained using the VIRTIS-M instrument, while \textbf{f} is a ground-based image from the SpeX instrument. Scale bars: 1,000 km. A version of this figure with cylindrical projections and corresponding grids is provided in Supplementary Fig.~\ref{figure:figS1}.}
\label{figure:night-clouds}
\end{figure}

\begin{figure}[h!]
\centerline{\includegraphics[width=20pc]{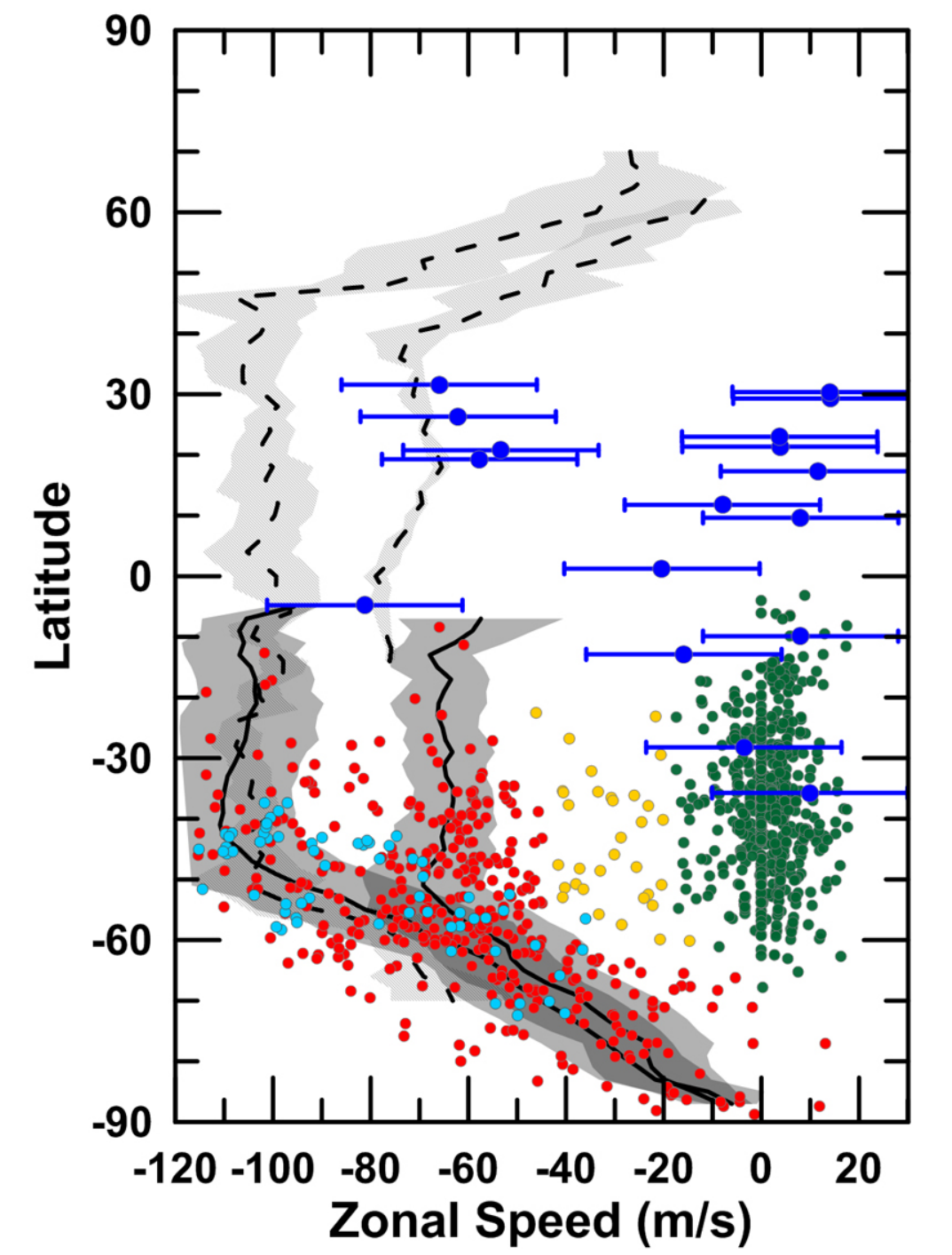}}
\caption{\textbf{Zonal velocities of thermal features on Venus's nightside.} Velocities for wavy and other patterns are represented by green and red dots, respectively. Fast filaments and shear-like features are represented by cyan dots. Eventual cases of extremely slow tracers without wavy morphologies are represented by yellow dots. Tracers in the SpeX images are represented by blue circles with their measurement error bars. Individual measurement errors with VIRTIS-M ranged from around 5 m s$^{-1}$ within pole-to-mid-latitudes to 15 m s$^{-1}$ at equatorial latitudes. Meridional profiles of zonal winds at dayside levels of the cloud tops ($\sim$70 km) and upper clouds' deck ($\sim$60 km) during the Galileo flyby \citep{Peralta2007b} and VEx missions \citep{Hueso2015} are represented by dashed and continuous black lines with shaded areas showing the dispersion of the measurements.}
\label{figure:zonal-velocities}
\end{figure}

\null
Nightside thermal emission features from the upper cloud level exhibit important day-to-day changes, revealing morphologies unseen in dayside ultraviolet images \citep{Titov2012} (Fig.~\ref{figure:night-clouds} and Supplementary Fig.~\ref{figure:figS1}). We tracked night features using pairs of VIRTIS-M images separated in time by 20 to 120 min --the 120 min interval also being used for SpeX observations-- to obtain 1,060 velocity measurements. Three types of features and motions were found: (1) wavy patterns, (2) patchy or irregular patterns and (3) filaments or shear-like patterns. Their motions (Fig.~\ref{figure:zonal-velocities}) ranged within the fast zonal winds (-110 m s$^{-1}$) found for dayside cloud tops ($\sim$70 km altitude) and the slower winds (-60 m s$^{-1}$) a few kilometers below ($\sim$60--65 km altitude) found in the Galileo \citep{Peralta2007b} and VIRTIS data \citep{Hueso2015}. Large-scale wavy features were the most abundant on the nightside (Fig.~\ref{figure:night-clouds}\textbf{a-c}); they were frequently seen between the equator and 60$^{\circ}$S and exhibited almost stationary motions with velocities of -10 to +10 m s$^{-1}$. The wave trains had wavelengths of around 100--250 km, were oriented $\pm$45$^{\circ}$ relative to the parallels, had packet lengths of around 1,000 km and extended 1,000--3,000 km. Their properties differed from the dayside gravity waves observed at similar altitudes \citep{Peralta2008,Piccialli2014}. Bright stretched filaments and shear-like features were less frequent (Fig.~\ref{figure:night-clouds}\textbf{d,e}). Well-contrasted bright and dark filaments from 40$^{\circ}$S to 70$^{\circ}$S with small features inside them and shear-like features at their edges all moved at velocities consistent with those found at the cloud tops during the dayside \citep{Kouyama2013,Khatuntsev2013,Hueso2015} (Fig.~\ref{figure:zonal-velocities}). The filaments were particularly interesting since they could be distinguished in their spectral signature by a distinct peak at 4.18 $\mathrm{\mu m}$ plausibly related to the presence of a stronger temperature inversion near the cloud tops (Supplementary Fig.~\ref{figure:figS2}). Most patchy features resembling cloud morphologies were between 10$^{\circ}$S and 60$^{\circ}$S and moved with variable velocities between -100 and -50 m s$^{-1}$, with a significant portion moving at about -60 m s$^{-1}$. In at least eight VEx orbits, very slow patchy features (slower than -40 m s$^{-1}$) were identified (see Supplementary Videos 15--21). Cloud features in the images from SpeX (Fig.~\ref{figure:night-clouds}\textbf{f}) also exhibited fast or slow motions in different areas of the images at low latitudes of 40$^{\circ}$N--40$^{\circ}$S, but the spatial resolution did not allow characterization of the tracers in terms of morphologies. The limited data distribution did not reveal clear dependence between velocities and local time. Poleward of 60$^{\circ}$S, the measured velocities agreed with those previously determined on the polar vortex \citep{Luz2011,Garate-Lopez2013}.\\

\begin{figure}[!tp]
\includegraphics[width=18pc]{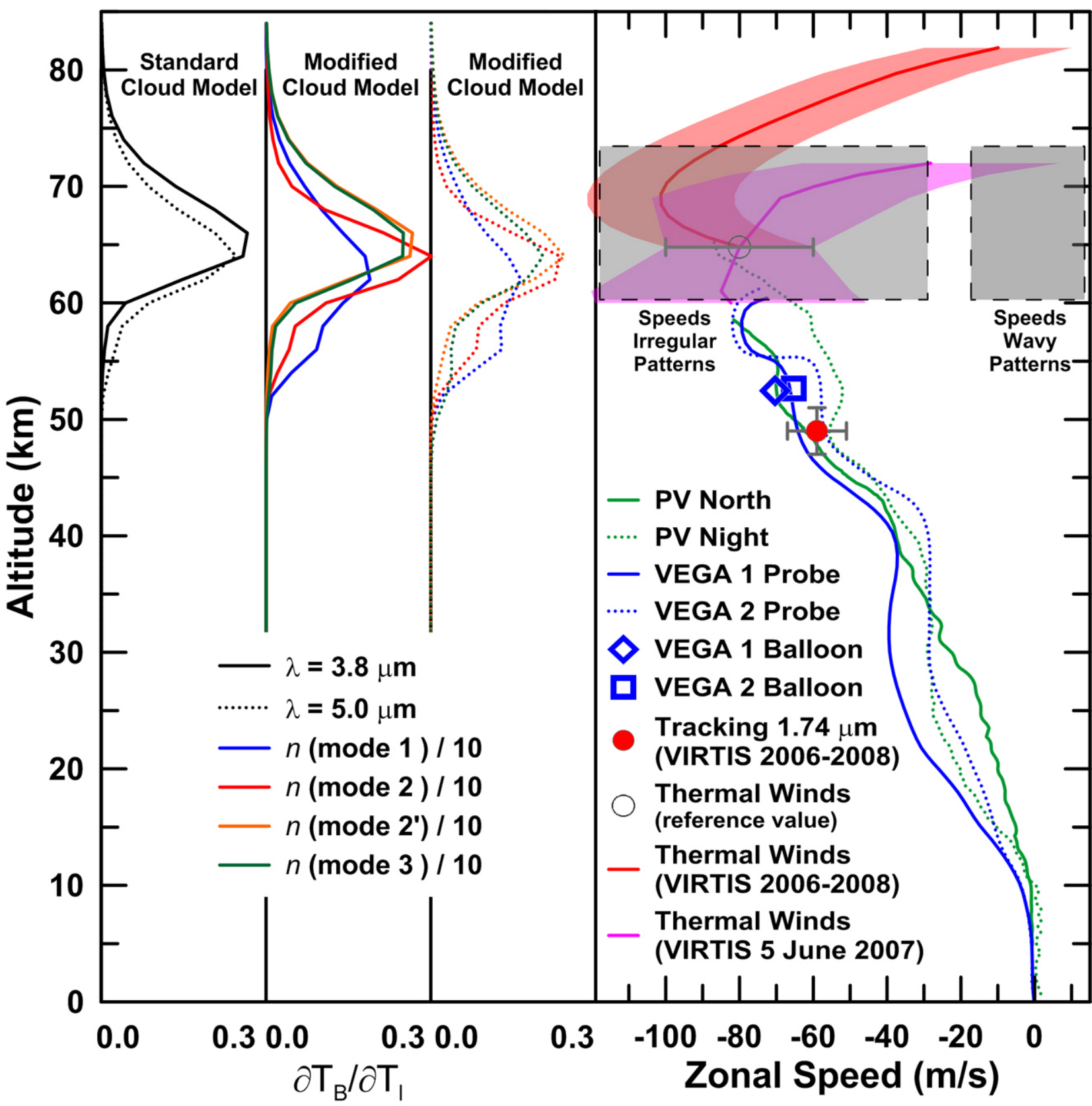}
\caption{\textbf{Vertical sensitivity at wavelengths of 3.8 and 5.0 $\mathrm{\mu m}$ and vertical profiles of zonal winds. a}, Jacobian representatives of the sensitivity of the models to different altitudes are displayed as continuous (3.8 $\mathrm{\mu m}$) and dotted lines (5.0 $\mathrm{\mu m}$). Black represents the nominal model, while colours represent non-nominal configurations with reduced number densities for each of the particle size modes. \textbf{b}, The velocity ranges for irregular and wavy features (this study) are displayed as rectangular areas and compared with relevant wind measurements from previous studies. \textit{In situ} zonal winds from Pioneer Venus (PV) North and Night probes \citep{Gierasch1997,Counselman1980} (entry locations 59$^{\circ}$N and 03:36 solar local time and 29$^{\circ}$S and 00:30) are shown by continuous and dotted green lines, while the VEGA probes \citep{Moroz1997} (8$^{\circ}$N and 00:30 and 7$^{\circ}$S and 01:48) are shown by continuous and dotted blue lines and the VEGA balloons are shown by a diamond and a square. The cloud-tracked velocities for the lower clouds \citep{Hueso2012} at around 48 km were averaged for 20--60$^{\circ}$S and local times 22:00--02:00, and they are displayed by a red dot. The zonal winds predicted by the thermal wind equation were calculated using temperatures retrieved from VIRTIS-M data and a reference value of -80 m s$^{-1}$ at around 65 km based on \textit{in situ} measurements using Pioneer Venus probes (empty circle). These thermal winds are presented for two cases: temperatures averaged for the years 2006--2008 \citep{Grassi2014} and the meridional gradient calculated for 30--60$^{\circ}$S and local times 23:00--01:00 (red line with an orange uncertainty region) and the instantaneous state during the Messenger's flyby \citep{Peralta2017b} on 5 June 2007, calculated for the intervals 00--30$^{\circ}$S and 19:00--20:00 (magenta line with pink uncertainty region).}
\label{figure:wfunction-zwind}
\end{figure}

\null
Nightside radiation at 3.8 and 5.0 $\mathrm{\mu m}$ was attributed to thermal emission from the middle and upper clouds \citep{GarciaMunoz2013,Carlson1991}. We reassessed previous altitude estimations using two radiative transfer models previously validated for Venus conditions \citep{GarciaMunoz2013,Lee2015} (see Methods). The models adopt a standard description of the Venus cloud particles (a sulfuric acid solution in water of 75\% concentration by weight) distributed vertically with four size modes with different number densities \citep{Crisp1986}. We used the vertical temperature profile at 45$^{\circ}$ latitude from the Venus International Reference Atmosphere \citep{Seiff1985}. Our calculations show that radiation at both wavelengths was sensitive to a range of altitudes between 60 and 72 km (Fig.~\ref{figure:wfunction-zwind}). Thinner clouds occurred occasionally in the Venus atmosphere \citep{Arney2014}, which may have resulted in larger contributions to the measured thermal emission from lower altitudes. To test such a scenario, four additional descriptions of the thermal opacity were explored reducing by a factor of ten the number density of a size mode, while leaving the other size modes unmodified. The calculations show that even in conditions of thinner clouds the bulk of radiation at 3.8 and 5.0 $\mathrm{\mu m}$ originated at altitudes above 50 km (Fig.~\ref{figure:wfunction-zwind}). The radiative transfer modeling provides confidence in that the motions under investigation occurred at the upper cloud level in the altitude range 60--70 km. The right panel of Fig.~\ref{figure:wfunction-zwind} compares our cloud-tracked zonal velocities (averaged between 20 and 60$^{\circ}$S and between 22:00 and 02:00) with the \textit{in situ} nocturnal measurements provided by the Pioneer Venus and VEGA probes/balloons \citep{Gierasch1997,Counselman1980,Moroz1997}, cloud-tracked winds from the lower clouds \citep{Hueso2012} and zonal winds predicted by the thermal wind equation using temperatures retrieved from VIRTIS-M data averaged for the years 2006--2008 \citep{Grassi2014} and during Messenger's flyby \citep{Peralta2017b} in 2007.\\

\null
The ensemble of nightside wind measurements could not be interpreted as a single mean wind profile with added variability and noise as was the case for the wind variability found on the dayside at altitudes of 60--70 km \citep{Hueso2015}. Instead, we identified several types of moving features. First, the best-contrasted features were shear-like and bright filaments, the bright filaments resembling the spiral patterns of cloud systems in the dayside images \citep{Titov2012}. These were the least frequent of the night features and moved with zonal velocities compatible with the circulation of the dayside cloud tops (cyan dots in Fig.~\ref{figure:zonal-velocities}). Second, patterns of nearly stationary waves were frequent and unambiguously observed (green dots in Fig.~\ref{figure:zonal-velocities}; see Supplementary Videos 01--14). Third, and most puzzlingly, the variability in the motions of the cloud-like features (red and yellow dots in Fig.~\ref{figure:zonal-velocities}) was subject to multiple non-exclusive interpretations: (1) features with small contrasts (2--5\% in radiance) could have originated at different altitudes (60--72 km) and their velocity variability could be a consequence of the vertical wind shear; (2) manifestation of wave motions could represent phase speed measurements instead of real fluid motion; and (3) part of the variability could have been caused by local real decelerations of the flow due to the drag of the waves. For instance, diurnal tides \citep{Limaye2007} could produce a nocturnal deceleration of up to 30 m s$^{-1}$ on the global winds in some latitudes and local times. At altitudes of 65--80 km, the radiative time constant was around 0.8--1.2 days \citep{Crisp1986} and circulation changes forced by solar radiation were expected to occur from day to night since the atmospheric superrotation had a period around 4 days. The magnitude of these changes close to the cloud tops remains to be quantified but could be constrained by the fact that dayside ultraviolet images show recurrent large cloud patterns appearing distorted after a full rotation above the planet at time scales of around 4 days \citep{Titov2012}. Thermal winds on the nightside seem to suggest that the mean zonal flow reaches it maximum velocity at altitudes deeper than on the dayside \citep{Peralta2017b}. If confirmed, the strong decrease in the thermal winds starting above the cloud tops (Fig.~\ref{figure:wfunction-zwind}) would support the hypothesis that the detected cases of extremely slow motions (yellow dots in Fig.~\ref{figure:zonal-velocities}) may correspond to real slow winds. However, caution must be taken since the calculations of the thermal winds assumed a zonally axisymmetric flow and the temperatures retrieved from VIRTIS-M were dependent on the cloud properties \citep{Grassi2014}.\\

\begin{figure}[h!]
\centerline{\includegraphics[width=20pc]{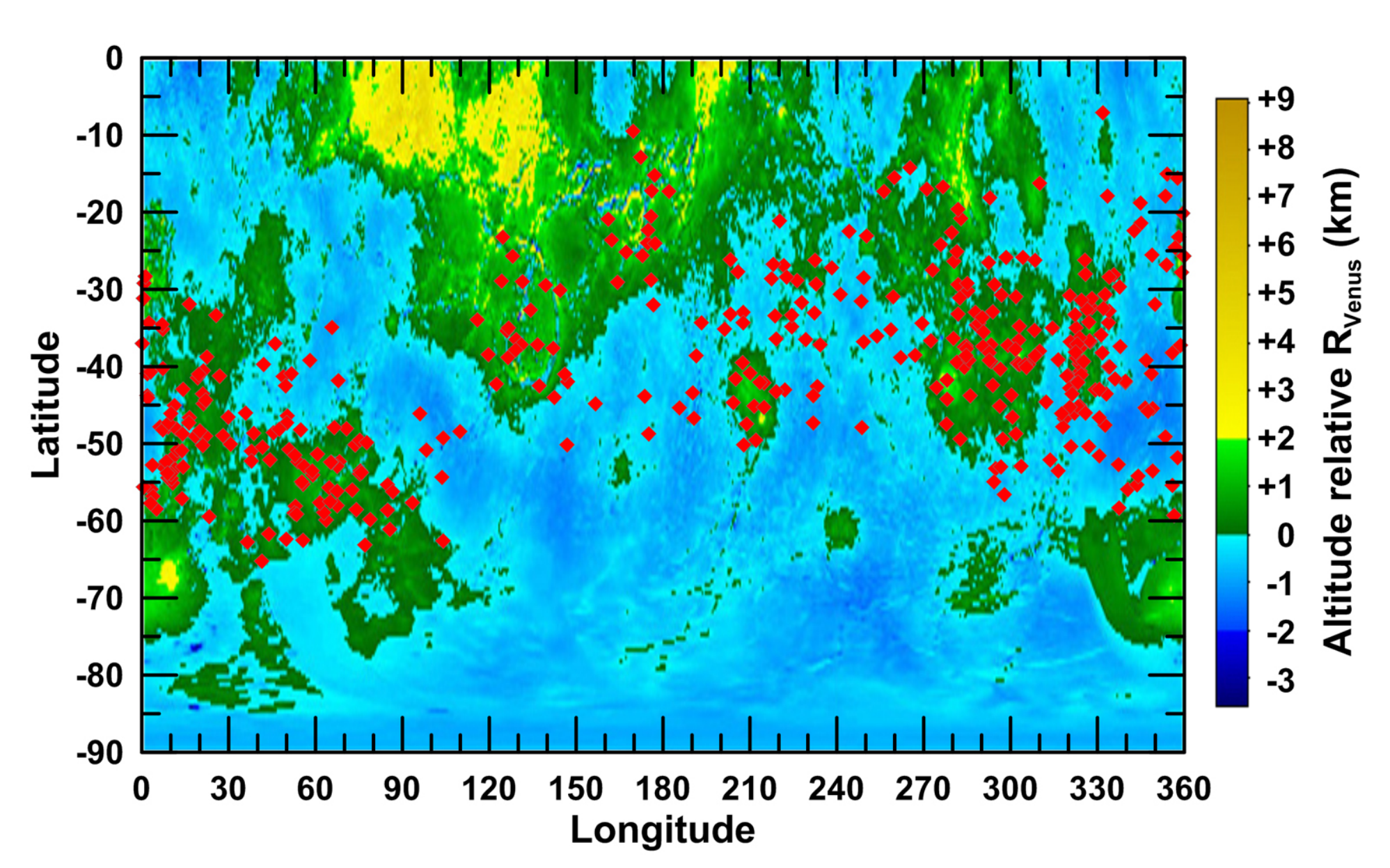}}
\caption{\textbf{Geographical distribution of the stationary waves found on the southern hemisphere with the VIRTIS-M images taken at 3.8 and 5.0 $\mathrm{\mu m}$.} A total of 408 stationary and quasi-stationary wavy patterns apparent in the thermal emission of the Venus nightside upper clouds are displayed (red diamonds) and compared with a cylindrical map of surface elevation for the southern hemisphere of Venus. Values of altitude are given in km relative to the mean planetary radius. The wave features concentrate around the areas of highest altitude, especially over Imdr, Themis and Alpha Regiones, and Lada Terra (see Supplementary Figure \ref{figure:figS6}\textbf{c}). The geographical distribution of the cloud features is displayed in Supplementary Fig.~\ref{figure:figS3}, which shows that the absence of waves over the southernmost area below 60$^{\circ}$S and between 120 and 270$^{\circ}$ is real, while the lack of positive identifications over Aphrodite Terra (about 60$^{\circ}$--150$^{\circ}$ longitude at the equator) is caused by the poor coverage of lower latitudes by the VIRTIS-M observations (see Supplementary Fig.~\ref{figure:figS3}).}
\label{figure:orographic-waves}
\end{figure}

\null
The nearly stationary waves in the nightside VIRTIS-M images are abundant. A global stationary bow identified on the dayside of Venus with Akatsuki thermal data \citep{Fukuhara2017} and indications of geographical dependence for the winds of the cloud tops \citep{Bertaux2016} and mesoscale waves \citep{Piccialli2014} suggest that airstreams may be triggering stationary waves up to the cloud tops over Aphrodite Terra and other high-elevation regions \citep{Fukuhara2017}. Figure \ref{figure:orographic-waves} displays the geographical location of the stationary waves from this study, demonstrating a high correlation with the main elevations of the southern hemisphere and proving that orographic waves are a quasi-permanent feature of the Venus atmosphere. This is a surprising result, not only because the formation and propagation of such waves is difficult considering the weak winds reported for the surface and the extended neutral stability of the lower atmosphere \citep{Gierasch1997,Fukuhara2017}, but also because the southern hemisphere is practically a planar surface with elevations rarely higher than 2 km ($\sim$3 km relative to the lowest planitia). Against expectations, stationary waves are not apparent in the simultaneous VIRTIS-M images taken at 1.74 and 2.3 $\mathrm{\mu m}$, which sense the deeper clouds at around 50 km above the surface and below the upper clouds where non-stationary mesoscale waves usually occur \citep{Peralta2008}. Scatter plots and principal component analysis (see Supplementary Figs.~\ref{figure:figS4} and \ref{figure:figS5}) reveal the lack of correlation between the morphologies in the 3.8 and 5.0 $\mathrm{\mu m}$ VIRTIS-M images and simultaneous images at the spectral bands sensing the surface thermal emission and the lower clouds at 1.02 and 1.74 $\mathrm{\mu m}$, respectively.\\

\null
These wave patterns can be interpreted as gravity waves supported by the positive static stability above 60 km \citep{Tellmann2012} and their position in a wave dispersion diagram \citep{Peralta2014a} (Supplementary Fig.~\ref{figure:figS6}\textbf{a}). To constrain the vertical wavelengths associated with the stationary waves, up to 63 thermal profiles from the Venus Express Radio Science Experiment (VeRa) were examined at the geographical locations displaying the highest concentration of waves (Supplementary Fig.~\ref{figure:figS6}\textbf{c}), resulting in vertical wavelengths ranging from 4 to 17 km (although the vertical coverage for VeRa temperature profiles rarely exceeds 20 km), while the Brunt-V\"{a}is\"{a}l\"{a} frequency ranged from 15$\times$10$^{-3}$ to 22$\times$10$^{-3}$ s$^{-1}$ between 60 and 70 km in height. These mesoscale stationary waves induce brightness temperature contrasts of about 5$\pm$3 K, which is similar to the 4--6 K associated with the planetary-scale bow-shape thermal feature discovered by Akatsuki \citep{Fukuhara2017}. The properties of the observed thermal waves notably differ from those of the waves seen in the daytime clouds \citep{Peralta2008,Piccialli2014}, which suggests a different effect in the atmosphere. We cannot rule out that the waves may lead to an effective wave drag on the mean zonal wind through the Eliassen-Palm flux mechanism \citep{Andrews1987}.\\

\null
The simultaneous presence of fast and slow motions and recurrent nearly stationary waves controlled by the orography on the nocturnal upper clouds of Venus will provide constraints to Venus general circulation models, which currently do not predict such phenomena. The Japan Aerospace Exploration Agency's Akatsuki \citep{Nakamura2016} mission to Venus is currently providing complementary observations that will help to further elucidate the features of the night- and dayside upper clouds of Venus.\\

\section*{Methods}

\textbf{Image processing and cloud tracking on VIRTIS-M images.} The absolute local contrast of low-latitude features in single wavelength VIRTIS-M thermal images at 3.8 or 5.0 $\mathrm{\mu m}$ is generally only of a few percent. To improve the signal-to-noise ratio and visibility of the tracked features, we stacked several consecutive wavelengths representative of the same altitude layers \citep{GarciaMunoz2013}. The spatial resolution of the images ranged nominally from about 10 km pixel$^{-1}$ for VIRTIS-M to about 200 km pixel$^{-1}$ for the SpeX \citep{Rayner2003} observations. However, the SpeX images were acquired with an atmospheric seeing estimated to be about 1 arcsec that limited the effective spatial resolution of individual frames to about 400 km pixel$^{-1}$. Super-resolution techniques combining multiple frames \citep{Farsiu2004,Mendikoa2016,Sanchez-Lavega2016} allowed us to improve over the atmospheric seeing and approach the diffraction limit (which is about 125 km at 5.0 $\mathrm{\mu m}$ for the IRTF). Here, we stacked several frames to subpixel accuracy after increasing the images size by a factor of three using a bilinear interpolation algorithm. After analysis of the stacked images, the effective spatial resolution was estimated to be about 200 km pixel$^{-1}$. VIRTIS-M and SpeX image pairs were projected onto cylindrical and/or polar projections with a spatial resolution similar to the worst of the two images of each pair, although in some cases a value of 0.1 deg pixel$^{-1}$ was considered for the coordinates of latitude and longitude, depending on the latitudes to be studied. Contrast of faint features was increased using high filtering based on the unsharp-mask technique. Alternatively, we applied a convolution with a weak form of a directional kernel previously used in the analysis of VIRTIS-M dayside images \citep{Hueso2015}. Cloud tracking was performed by comparing images separated by 20 to 120 min. As VIRTIS-M is a scanning spectrometer \citep{Piccioni2007ESA}, each horizontal image in a given image cube is obtained at a different time. In the images separated by 30 min or less, the specific time of each horizontal line was taken into account to produce an accurate measurement. The cloud-tracked measurements presented here were performed visually and independently by three of the authors. Two of us used zoomed versions of the images with Planetary Laboratory for Image Analysis software \citep{Hueso2010}, while the other used custom-made tools. Automatic cloud correlation measurements were also tested in some orbits, yielding results coherent with those presented in Fig.~\ref{figure:zonal-velocities}. Nevertheless, these were generally noisier and they were finally discarded. Individual errors for wind measurements using VIRTIS-M data were variable: from 15 m s$^{-1}$ for the very few tracers from images separated by short time intervals ($\sim$20 min) to about 5 m s$^{-1}$ in most other cases at mid- to polar latitudes, and up to 15 m s$^{-1}$ at equatorial latitudes. Individual errors for wind measurements taken from the SpeX images were in the order of 20--25 m s$^{-1}$.\\

\null
Radiative transfer modeling. We analyzed the altitudes probed with the 3.8 and 5.0 $\mathrm{\mu m}$ images using two different radiative transfer models.

\begin{itemize}

	\item We used the radiative transfer model from \textit{Garc{\'i}a-Mu{\~n}oz et al.} \cite{GarciaMunoz2013}, which solves the radiative transfer equation within the atmosphere with a line-by-line treatment of absorption and scattering. The optical properties for both the gases and aerosols were pre-calculated and tabulated. The radiative problem was solved in a plane parallel, stratified atmosphere, and considered multiple scattering. The clouds were prescribed by the number densities of each particle size mode and the wavelength-dependent values of their cross sections. In our nominal scenario, we prescribed number densities based on \textit{Crisp} \cite{Crisp1986} and particle size distributions based on \textit{Wilson et al.} \cite{Wilson2008}. We included the four particle size modes (so-termed 1, 2, 2' and 3) that are thought to make up the Venus atmosphere and considered altitudes from 30 to 100 km. All the size modes were assumed to have the same composition (75\% sulfuric acid by mass). Other non-nominal scenarios were considered. In each of them, we reduced the number density of one particle size mode (and only one) by a factor of ten (see Fig.~\ref{figure:wfunction-zwind}). These scenarios were introduced to consider the effect of clouds locally thinner than in the standard configuration \citep{Arney2014}.

	\item We used the radiative transfer model from \textit{Lee et al.} \cite{Lee2015} to constrain the lowest atmospheric levels that may contribute to 3.8 and 5.0 $\mathrm{\mu m}$ images. The model performs a line-by-line calculation for the entire Venus atmosphere
(0--100 km) considering various cloud structures and gaseous absorptions including collision-induced absorption \citep{Moskalenko1979} and using the cloud model described by \textit{Crisp} \cite{Crisp1986}: lower cloud (48--50 km), middle cloud (51--60 km) and upper cloud ($>$60 km). We explored the 3.3--5.2 $\mathrm{\mu m}$ (1,900--3,000 cm$^{-1}$) wavelength ranges investigating extreme scenarios including: (1) an atmosphere of pure CO$_2$ without collision-induced absorption, (2) a pure CO$_2$ atmosphere with collision-induced absorption and (3) a more realistic atmosphere including contributions from the rest of the known gases (N$_2$, H$_2$O, CO, SO$_2$, OCS, H$_2$S, HF and HCl). We used the temperature profile from the Venus International Reference Atmosphere \citep{Seiff1985} and considered emission angles of 0$^{\circ}$ and 45$^{\circ}$. As in the previous case, we considered variations in cloud particle numbers by a factor of 0.1 for each of the three cloud layers. Three situations were explored: (1) perturbation only for the middle cloud opacity, (2) perturbation only for the middle cloud opacity with 'thinner' upper cloud and (3) perturbation in the lower clouds with 'thinner' upper and middle clouds. While in the first situation (nominal upper and lower clouds) changes in the outgoing radiance due to the middle clouds can be regarded as negligible, noticeable influences were seen for the other two situations, which have in common upper clouds thinner than nominal. The total cloud opacity was 30--31 as calculated at 1 $\mathrm{\mu m}$, which is consistent with a recent study \citep{Haus2014}.

\end{itemize}

\null
For both models, we concluded that no important contribution could be originated from altitudes below a height of 40 km. Based on this result, we ruled out influences from the surface and the deep troposphere of Venus where prevailing winds would be more consistent with the slow features.\\

\null
\textbf{Data availability.} The full VIRTIS-M dataset of images that support the findings of this study are available from the public \href{https://www.cosmos.esa.int/web/psa/ftp-browser}{ESA repository}. The IRTF/SpeX images and VeRa temperature profiles are available from the authors on reasonable request (see Author contributions for specific datasets). The data that support the plots and other findings of this study are available from the corresponding author upon
reasonable request.

\section*{Acknowledgements}
J.P. acknowledges the Japan Aerospace Exploration Agency's International Top Young Fellowship. R.H. and A.S.-L. were supported by the Spanish project MINECO/FEDER, UE (AYA2015-65041-P) and Grupos Gobierno Vasco (IT-765-13). T.M.S. was supported by a Grant-in-Aid for the Japan Society for the Promotion of Science Fellows. The IRTF/SpeX observations were supported by the Japan Society for the Promotion of Science (KAKENHI 15K17767). T.K., T.M.S. and H.S. were visiting astronomers at the IRTF, which is operated by the University of Hawaii under contract NNH14CK55B with the National Aeronautics and Space Administration, and acknowledge M. S. Connelley (Institute for Astronomy, University of Hawaii) for support in the observations. We also thank the Agenzia Spaziale Italiana and the Centre National d'{\'E}tudes Spatiales for supporting the VIRTIS/VEx experiment.

\section*{Author contributions}
J.P. explored, selected and processed the data from the VIRTIS dataset, wrote the manuscript and produced Figs.~1--4 and Supplementary Figs.~1, 3 and 6. R.H. evaluated the signal-to-noise ratios of the images, the correspondence with deeper levels, performed the PCA analysis and produced Supplementary Figs.~4 and 5. A.S.-L. suggested the scheme for the manuscript, as well as some of the figures to be included, and chaired the discussion of the results. J.P., R.H. and A.S.-L. measured the cloud motions from VIRTIS and J.P. and T.K. measured those from SpeX. Y.-J.L. and A.G.M. coordinated, designed and carried out the sensitivity analyses at the wavelengths of interest. Y.-J.L. studied the spectral features of the filaments and produced Supplementary Fig.~6. T.K., T.M.S. and H.S. obtained, reduced, corrected and navigated the SpeX images. S.T. measured the atmospheric stability and vertical wavelengths in the VEx and VeRa radio-occultation data. All the authors discussed the results and commented on the manuscript.

\section*{Additional information}
\textbf{Supplementary information} is \href{https://www.nature.com/articles/s41550-017-0187#supplementary-information}{available for this paper}.\\

\textbf{Reprints and permissions} information is available at \href{www.nature.com/reprints}{www.nature.com/reprints}.\\

\textbf{Correspondence and requests for materials} should be addressed to J.P.\\

\textbf{How to cite this article:} Peralta, J. \textit{et al.} Stationary waves and slowly moving features in the night upper clouds of Venus. \textit{Nat. Astron.} \textbf{1}, 0187 (2017).\\

\textbf{Publisher's note:} Springer Nature remains neutral with regard to jurisdictional claims in published maps and institutional affiliations.

\section*{Competing interests}
The authors declare no competing financial interests.



\bibliographystyle{naturemag}

\newpage

\onecolumn

\appendix

\setcounter{figure}{0}

\section*{Supplementary Figure 1}
\begin{figure}[h!]
\centerline{\includegraphics[width=35pc]{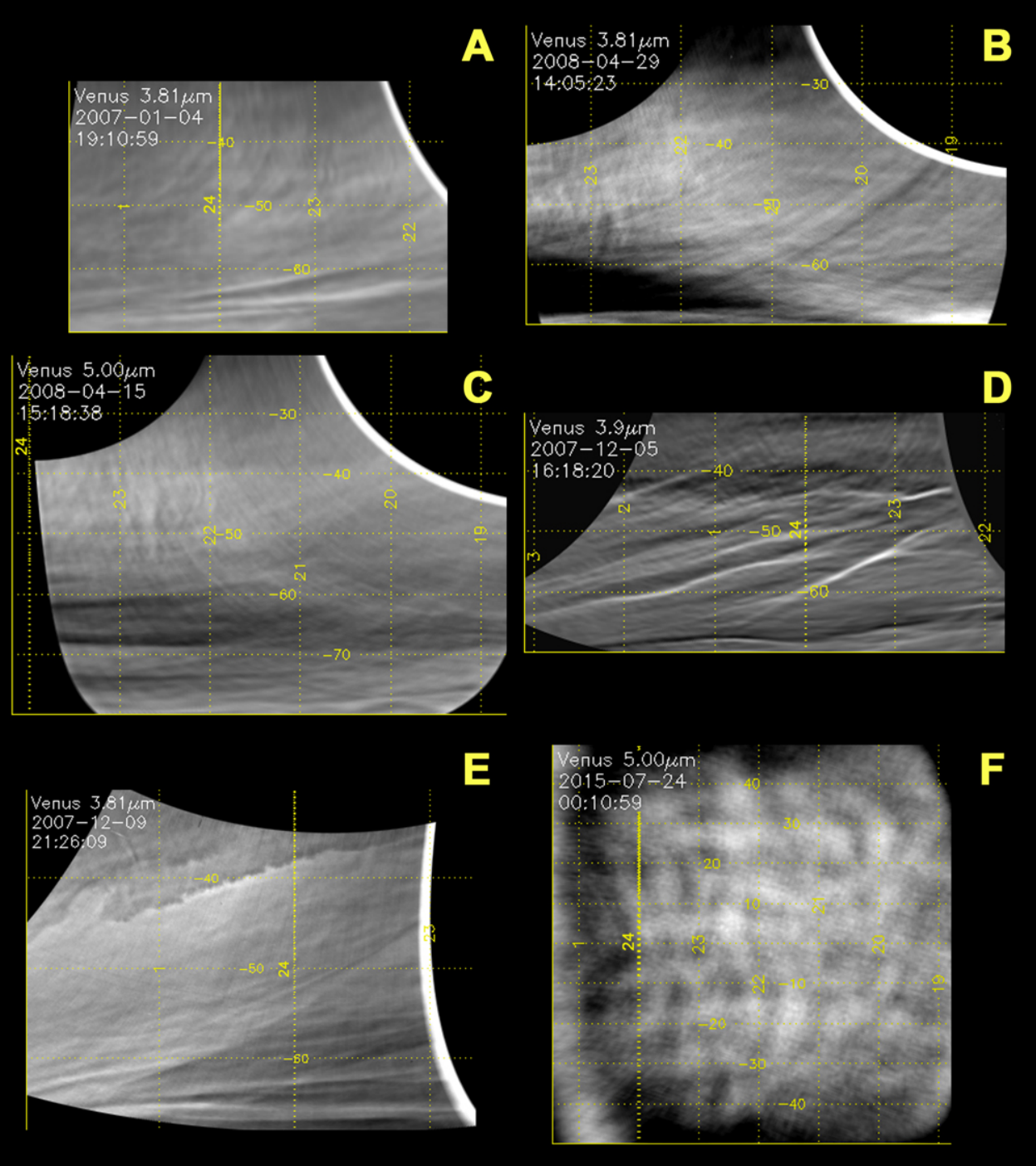}}
\caption{\textbf{Night side thermal features on the upper clouds of Venus.} (\textbf{A}) wavy patterns at 3.8 $\mathrm{\mu m}$; (\textbf{B}) long and broad stripes (probably waves) observed at 3.8 $\mathrm{\mu m}$; (C) further wavy patterns at 5.0 $\mathrm{\mu m}$; (\textbf{D}) bright filaments at 3.8 $\mathrm{\mu m}$; (\textbf{E}) shear-like structures from 3.8 $\mathrm{\mu m}$ images; (\textbf{F}) big-scale features observed in 5-$\mathrm{\mu m}$ images. Images A-E were obtained with the VEx/VIRTIS-M instrument, while F is a ground-based image from IRTF/SpeX. These images correspond to the cylindrical projections (latitude vs local time) of the panels displayed in Figure \ref{figure:night-clouds} of this work.}
\label{figure:figS1}
\end{figure}

\newpage

\section*{Supplementary Figure 2}
\begin{figure}[h!]
\centerline{\includegraphics[width=35pc]{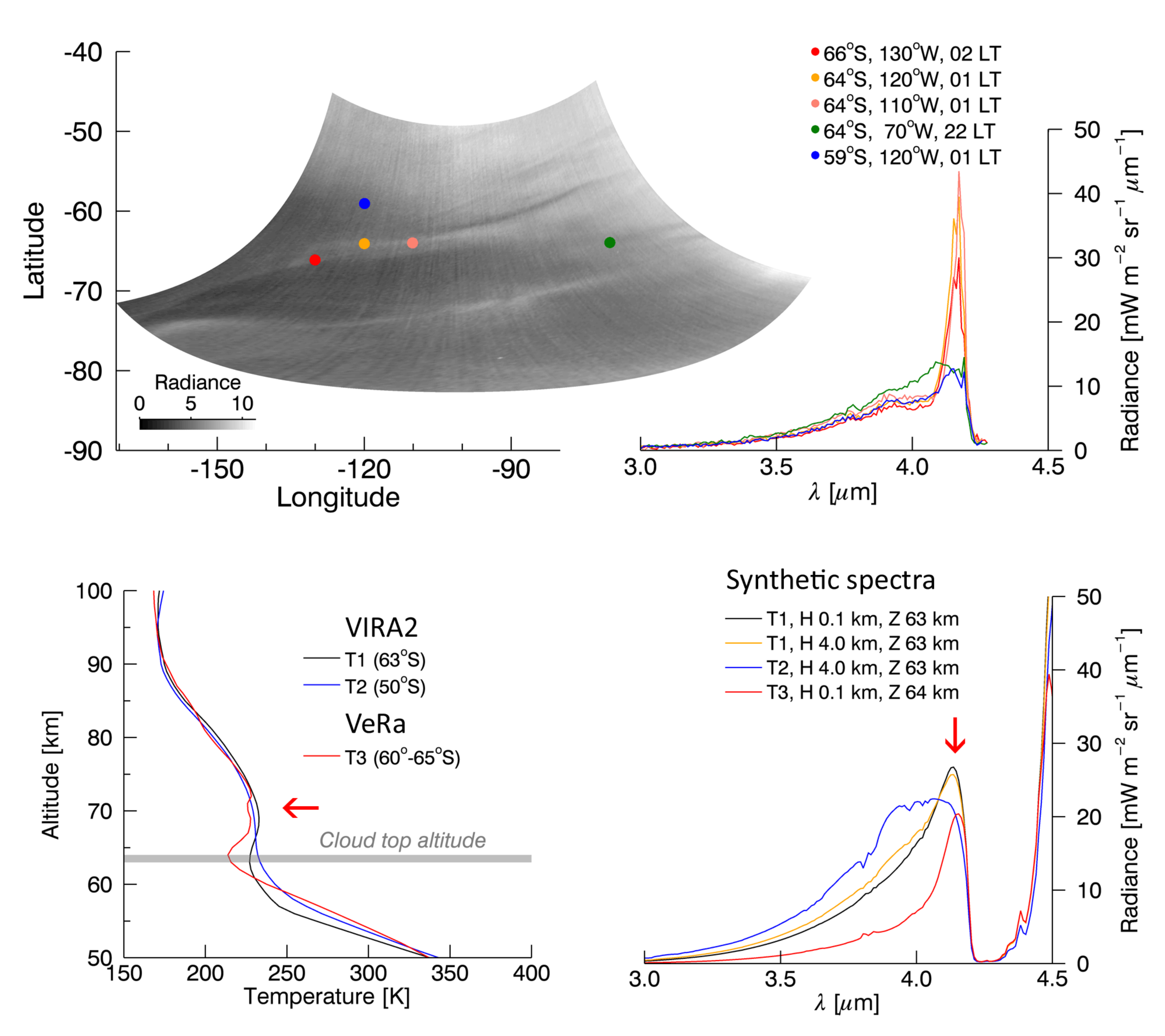}}
\caption{\textbf{Example of a filament at 3.8 $\mathrm{\mu m}$, and its spectral features.} Upper panel shows a cylindrically projected image taken by VIRTIS-M (VI0345\_00) and spectra at selected locations. The filament extends from 66$^{\circ}$S 130$^{\circ}$W through 64$^{\circ}$S 120$^{\circ}$W to 64$^{\circ}$S 110$^{\circ}$W, presenting significant spectral peaks at 4.18 $\mathrm{\mu m}$. Simulated spectra are calculated using temperature profiles \citep{Lee2015,Moroz1997} shown in the lower left panel and the 63--64 km cloud top altitude. Only the temperature inversion case can mimic the observed spectral shape (red arrows), but the absolute radiance value of the peaks would require much stronger temperature inversion than that yet observed.}
\label{figure:figS2}
\end{figure}

\newpage

\section*{Supplementary Figure 3}
\begin{figure}[h!]
\centerline{\includegraphics[width=30pc]{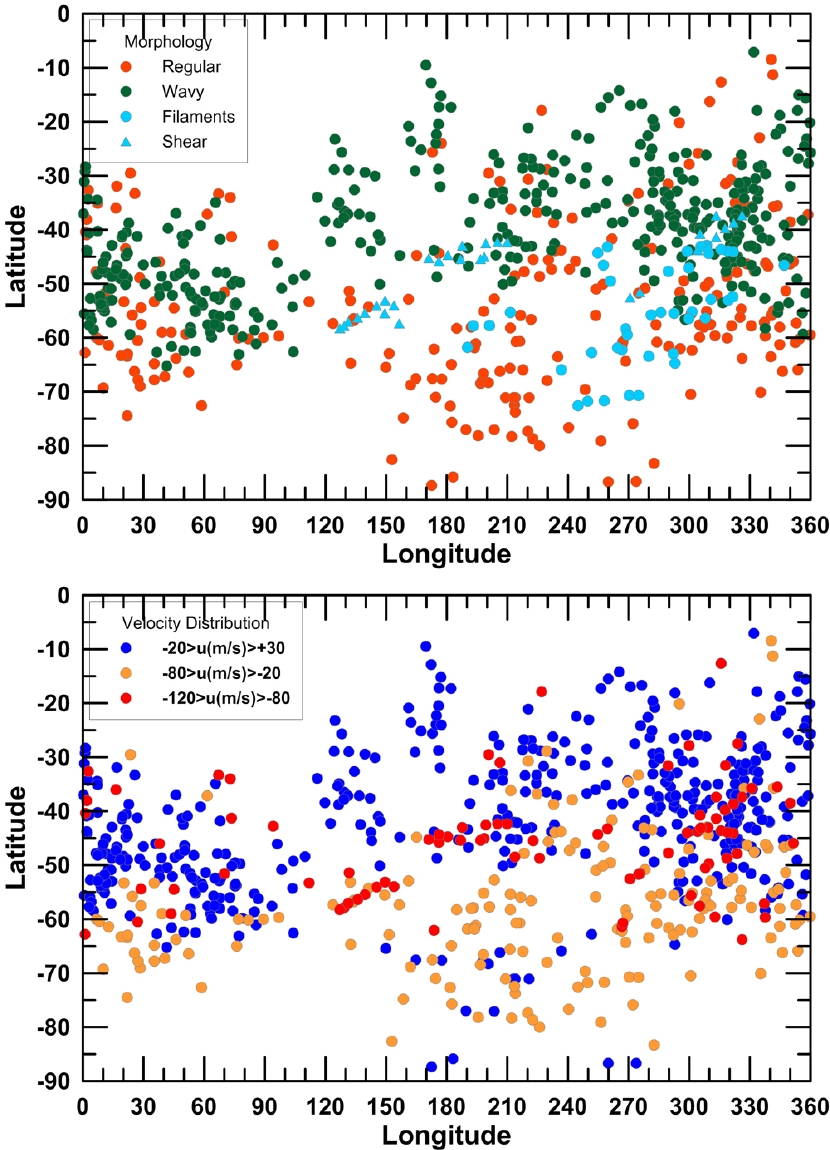}}
\caption{\textbf{Geographical Distribution of our motion measurements with VIRTIS-M images.} Above: location of the different thermal cloud morphologies. Below: location of our measurements in terms of ranges of speeds. This figure can be compared with the Figure \ref{figure:orographic-waves} in main manuscript to confirm whether the absence of wavy patterns in some areas is related or not to observational constrains of the VEx/VIRTIS-M observations.}
\label{figure:figS3}
\end{figure}

\newpage

\section*{Supplementary Figure 4}
\begin{figure}[h!]
\centerline{\includegraphics[width=12pc]{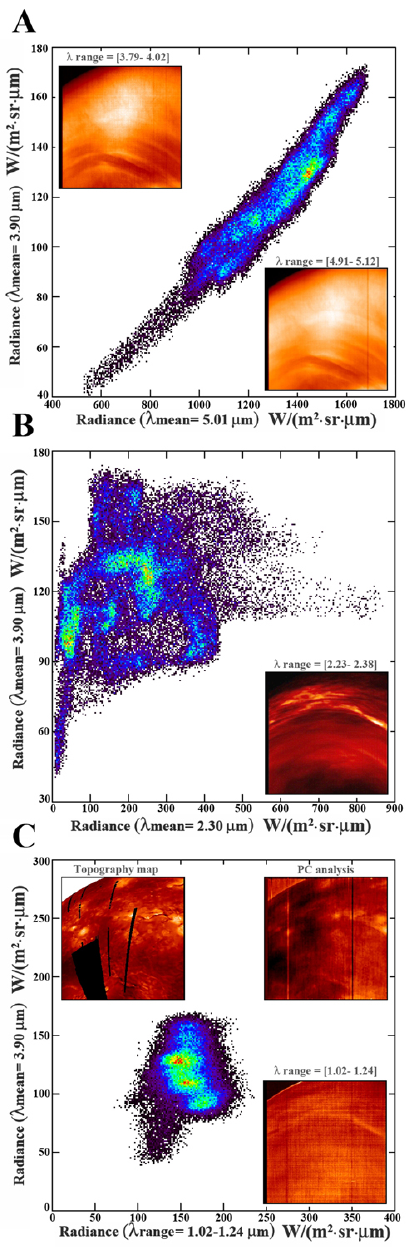}}
\caption{\textbf{Comparative morphologies and scatter plots at 3.90, 5.01, 2.30 and 1.02--1.24 $\mathrm{\mu m}$ for cube VIRTIS cube VI0824\_02.} (\textbf{A}) Scatter plot of radiances at 3.9 and 5.01 $\mathrm{\mu m}$ show a high degree of linear correlation which is enhanced by a visual comparison of images at these wavelengths showing essentially the same features with slightly different contrasts (insets);  (\textbf{B}) Scatter plot of radiances at 3.9 and 2.3 $\mathrm{\mu m}$ show uncorrelated radiances and a very different image (inset) to images at 3.9 and 5.01 $\mathrm{\mu m}$ (insets in A);  (\textbf{C}) Scatter plot of radiances at 1.02-1.24 $\mathrm{\mu m}$ and 3.9 $\mathrm{\mu m}$ show also uncorrelated radiances and a very different image (right-bottom inset in C) to images at 2.3, 3.9 and 5.01 $\mathrm{\mu m}$. All scatter plots are based in co-adding several close wavelengths to minimize pixel noise as indicated in the insets. Red and blue in the scatter plots represent higher and lower densities of pixels respectively. A Principal Component Analysis of the images in the 1.02--1.24 $\mathrm{\mu m}$ range results in a principal component image (upper right inset in C) which shows surface features also shown as an additional inset (upper left inset in C).}
\label{figure:figS4}
\end{figure}

\newpage

\section*{Supplementary Figure 5}
\begin{figure}[h!]
\centerline{\includegraphics[width=35pc]{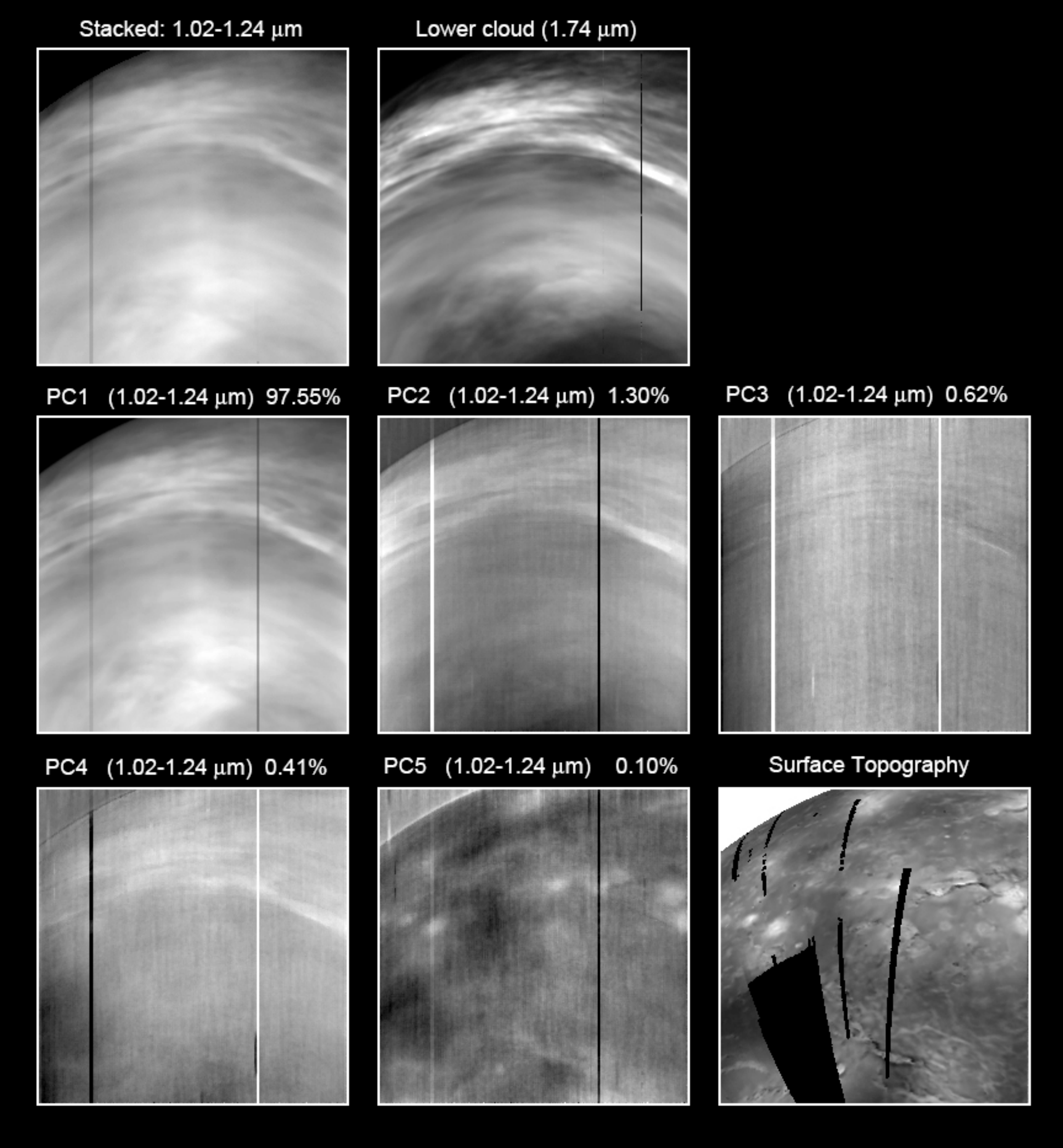}}
\caption{\textbf{Principal Component Analysis of the lower atmosphere contribution.} Information from the lower atmosphere can be obtained when examining spectral windows between 1.02 and 1.24 $\mathrm{\mu m}$ (upper row left panel) which are contaminated with information from the lower cloud as observed in 1.74 $\mathrm{\mu m}$ images (upper row central panel) with both images presenting a high degree of correlation. The spectral region 1.02--1.24 $\mathrm{\mu m}$ is analyzed in the next 5 panels (middle and bottom rows) using a Principal Component Analysis (PCA) that shows the lower cloud to contribute to a 97.55\% of the signal. PCA isolates different contributions related with the lower cloud and the surface contribution in the Principal Component 5 with a level of signal of 0.10\%. This PC5 is compared with the surface topography showing a high degree of correlation. This analysis corresponds to VIRTIS cube VI824\_02 but similar results are obtained in other cubes.}
\label{figure:figS5}
\end{figure}

\newpage

\section*{Supplementary Figure 6}
\begin{figure}[h!]
\centerline{\includegraphics[width=35pc]{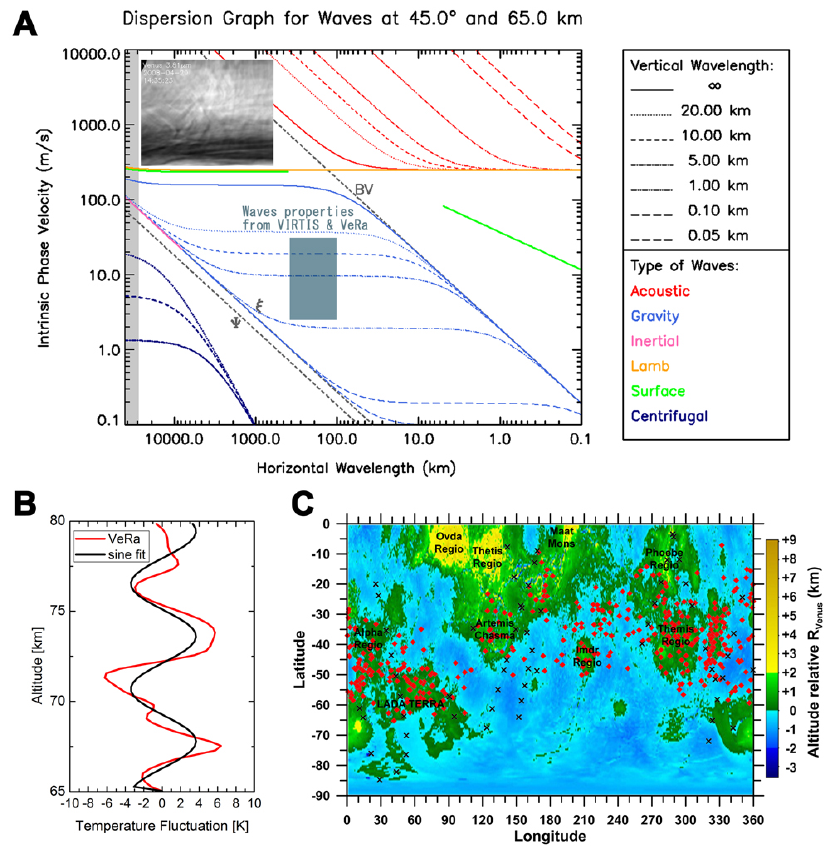}}
\caption{\textbf{Wave characterization of the slow wavy patterns found at the nightside upper clouds from images at 3.8 and 5.0 $\mathrm{\mu m}$.} Dispersion graph (panel \textbf{A}) for atmospheric waves on Venus at a latitude of 45$^{\circ}$ and altitude of 65 km. The theoretical solutions each type of waves are shown with different colors, while vertical wavelengths are marked with different line styles. The Brunt-V{\"a}is{\"a}l{\"a} frequency (BV), centrifugal frequency ($\Psi$), and centrifugal frequency modified by the meridional shear of the zonal wind ($\xi$) are marked with grey lines, assuming standard values from \textit{Peralta et al.} \cite{Peralta2014a}. The dark grey area mark the limits for maximum zonal wavelength allowed at this latitude. A blue rectangle represents the range of horizontal and vertical wavelengths of the stationary waves, with horizontal wavelengths measured in the VIRTIS-M images. The range of vertical wavelengths were calculated from a set of temperature profiles obtained with VEx/VeRa radio-occultation data \citep{Tellmann2012} (see example in panel \textbf{B}). Since most of the data from VeRa and VIRTIS-M was not acquired simultaneously, we chose temperature profiles (black crosses in panel \textbf{C}) on geographical locations close to the areas of highest abundance of stationary waves (red diamonds).}
\label{figure:figS6}
\end{figure}

\end{document}